\documentclass[12pt]{amsart}
\usepackage{amssymb}

\usepackage{float}
\usepackage{lmodern}
\usepackage{graphicx}

\theoremstyle{plain}

\numberwithin{equation}{section}

\begin{document}
\title{
Orthogonal Polynomials attached to coherent states for the symmetric
P\"{o}schl-Teller oscillator
}

\author{KHALID AHBLI$^{\ast}$, PATRICK KAYUPE KIKODIO$^{\flat}$ and ZOUHA\"{I}R MOUAYN$^{\natural}$}

\maketitle
\thanks{}

\begin{abstract}
We consider a one-parameter family of nonlinear coherent states by
replacing the factorial in coefficients $z^{n}/\sqrt{n!}$ of the canonical
coherent states by a specific generalized factorial $x_{n}^{\gamma }!$, $%
\gamma \geq 0.$ These states are superposition of eigenstates of the
Hamiltonian with a symmetric P\"{o}schl-Teller potential depending on a
parameter $\nu >1$. The associated Bargmann-type transform is defined for $%
\gamma =\nu $. Some results on the infinite square well potential are also
derived. For some different values of $\gamma ,$ we discuss two sets of
orthogonal polynomials that are naturally attached to these coherent states.
\end{abstract}

\section{Introduction}

Coherent \ states (CS) have attracted much attention in the recent decades.
\ They are a useful mathematical framework for dealing with the connection
between classical and quantum formalisms. \emph{Nonlinear} coherent states (NLCS) were build as extensions of the canonical CS of the harmonic oscillator and have become a tool of great importance in quantum optics in view of their perspective applications in the growing field of quantum technologies. see $\left[ 1\right] $ and references therein

In this paper, we replace the factorial $n!$ occurring in coefficients $%
z^{n}/\sqrt{n!}$ of the canonical CS by a specific generalized
factorial $x_{n}^{\gamma }!=x_{0}^{\gamma }x_{1}^{\gamma }\cdots x_{n}^{\gamma }$%
, where $x_{n}^{\gamma }$ is a sequence of positive numbers (given by $(\ref
{4.3})$ below) and $\gamma \in \left( 0,\infty \right) $ being a parameter.
The new coefficients are then used to consider a superposition of
eigenstates of the Hamiltonian with a symmetric P\"{o}schl-Teller (SPT$)$
potential depending on a parameter $\nu >1$. The obtained states constitute
a family of NLCS. For $\gamma =\nu $, we define the associated Bargmann-type
transform and we derive some results on the infinite square well potential.
Next, we proceed by a general method $\left[ 2\right] $ to discuss, for
different values of $\gamma ,$ two sets of orthogonal polynomials that are
naturally associated with these NLCS. One set of these polynomials is obtained from a symmetrization of the measure
which gives the resolution of the identity for the NLCS. The second set of polynomials
arises from the \textit{shift operators} [1-6]
attached to these NLCS.

The paper is organized as follows. In Section 2, we recall  NLCS
formalism we will be using. Section 3 is devoted to NLCS with a specific
sequence of positive numbers. \ In Section 4, these NLCS are attached to the
Hamiltonian with a symmetric P\"{o}schl-Teller potential. In Section 5, we
discuss, for some different values of the parameter $\gamma$, two sets of orthogonal polynomials that are associated with these NLCS. Section 6 is devoted to some remarks.

\section{Nonlinear coherent states}

In this section, we summarize the construction ([6], pp.146-151) of the
so-called deformed coherent states, also known as NLCS in quantum optics [7]. \\ For this, let us first recall the
series expansion definition of the canonical CS, which first
was due to Iwata [8]:
\begin{equation}
|z\rangle =(e^{z\bar{z}})^{-1/2}\sum\limits_{n=0}^{+\infty }\frac{\bar{z}^{n}%
}{\sqrt{n!}}|\varphi _{n}\rangle ,\quad z\in \mathbb{C},  \label{2.1}
\end{equation}
where the kets $|\varphi _{n}\rangle ,\,\,n=0,1,2,...,\infty $, are an
orthogonal basis in an arbitrary (complex, separable, infinite dimensional)
Hilbert space $\mathcal{H}$.\\ The related NLCS are defined as follows. Let $%
\{x_{n}\}_{n=0}^{\infty },\,\,x_{0}=0$, be an infinite sequence of positive
numbers. Let $\lim_{n\rightarrow +\infty }x_{n}=R^{2}$, where $R>0$ could be
finite or infinite, but not zero. We shall use the notation $%
x_{n}!=x_{1}x_{2}\cdots x_{n}$ and $x_{0}!=1$. For each $z\in \mathcal{D}$
some complex domain, a generalized version of $(\ref{2.1})$ can be defined as
\begin{equation}
|z\rangle =(\mathcal{N}(z\bar{z}))^{-1/2}\sum\limits_{n=0}^{+\infty }\frac{%
\bar{z}^{n}}{\sqrt{x_{n}!}}|\varphi _{n}\rangle ,\quad z\in \mathcal{D}
\label{2.2}
\end{equation}
where
\begin{equation}
\mathcal{N}(z\bar{z})=\sum\limits_{n=0}^{+\infty }\frac{|z|^{2n}}{x_{n}!}
\label{2.3}
\end{equation}
is a normalization factor chosen so that the vectors $|z\rangle $ are
normalized to one. These vectors $|z\rangle $ are well
defined for all $z$ for which the sum $(\ref{2.3})$ converges, i.e. $%
\mathcal{D}=\{z\in \mathbb{C},|z|<R\}$. We assume that there
exists a measure $d\nu $ on $\mathcal{D}$ ensuring the following resolution of the
identity
\begin{equation}
\int_{\mathcal{D}}|z\rangle \langle z|d\nu (z,\bar{z})=1_{\mathcal{H}}
\label{2.4}
\end{equation}
\\
Setting $d\nu (z,\bar{z})=\mathcal{N}(z\bar{z})d\eta (z,\bar{z})$, it
is easily seen that in order for $(\ref{2.4})$ to be satisfied, the measure $%
d\eta $ should be of the form
\begin{equation}
d\eta (z,\bar{z})=\frac{d\theta }{2\pi }d\lambda (\rho ),\quad z=\rho
e^{i\theta }  \label{2.5}
\end{equation}
where the measure $d\lambda $ solves the moment problem
\begin{equation}
\int_{0}^{R}\rho ^{2n}d\lambda (\rho )=x_{n}!,\quad n=0,1,2,...\, .
\label{2.6}
\end{equation}
In most of the practical situations, the support of the measure $d\eta $ is the whole domain $\mathcal{D%
}$, i.e., $d\lambda $ is supported on the entire interval $[0,R)$.
\newpage
To illustrate this formalism, we consider, as a first example, the infinite
sequence
\begin{equation}
x_{n}=n,\quad n=0,1,2,...\, ,
\end{equation}
so that $R=\infty $ and the problem stated in $(\ref{2.6})$ is the Stieljes
moment problem
\begin{equation}
\int_{0}^{+\infty }\rho ^{2n}d\lambda (\rho )=n!,\quad n=0,1,2,...\, .
\end{equation}
So that the appropriate measure is
\begin{equation}
d\lambda (\rho )=2e^{-\rho ^{2}}\rho d\rho ,\ \ 0\leq \rho <\infty .
\label{canonicalcoherent}
\end{equation}
In this case, we recover the canonical coherent states $(\ref{2.1})$. \\
A second example corresponds to the sequence of positive numbers
\begin{equation}
x_{n}=n\left( 2\sigma +n-1\right) ,\ \ \ n=0,1,2,3,... \, ,  \label{2.7}
\end{equation}
whith $2\sigma =1,2,3,...$ $,$ being a fixed parameter.
So that the moment problem is now
\begin{equation}
\int_{0}^{+\infty }\rho ^{2n}d\lambda (\rho )=n!(2\sigma )_{n}
\end{equation}
where $(a)_{n}=a(a+1)\cdots(a+n-1)$,$\ (a)_{0}=1$, is the shifted factorial.
The solution of this problem is
\begin{equation}
d\lambda (\rho )=\frac{2}{\pi }K_{2\sigma -1}(2\rho )\rho ^{2-2\sigma }d\rho
,\quad 0\leq \rho <+\infty ,
\end{equation}
where
\begin{equation}
K_{\tau }(x)=\frac{1}{2}\left( \frac{x}{2}\right) ^{\tau }\int_{0}^{+\infty
}\exp \left( -t-\frac{x^{2}}{4t}\right) \frac{dt}{t^{\tau +1}},\ \ \Re (x)>0,
\end{equation}
is the Macdonald function of order $\tau $ ([9], p.183). Here $R=\infty
$ and the associated coherent states are of Barut-Girardello type [10]:
\begin{equation}
|z,\sigma \rangle =\frac{|z|^{2\sigma -1}}{\sqrt{I_{2\sigma -1}(2|z|)}}%
\sum\limits_{n=0}^{+\infty }\frac{\bar{z}^{n}}{\sqrt{n!(2\sigma )_{n}}}%
|\varphi _{n}\rangle ,\quad z\in \mathbb{C},  \label{3.13}
\end{equation}
$I_{\tau }(.)$ being the modified Bessel function of the first kind and of
order $\tau $ ([9], p.172).

\section{NLCS with a specific sequence of positive numbers}

Here, we will be dealing with a one-parameter family of NLCS on the complex
plane, which interpolates between slightly modified canonical CS and a class of CS of Barut-Girardello type without
specifying the Hamiltonian system. Precisely, let $\gamma \in \lbrack
0,\infty )$ be a fixed parameter and let us define a set of NLCS associated
with the infinite sequence of positive numbers $x_{0}^{\gamma }=0,\
x_{1}^{\gamma }=\Gamma (2\gamma +1)$ and
\begin{equation}
x_{n}^{\gamma }:=\frac{n(n+\gamma )(n+2\gamma -1)}{n+\gamma -1},\ n=2,3,4,...%
\text{ },  \label{positivenumbers}
\end{equation}
by the superposition
\begin{equation}
\left| z;\gamma \right\rangle :=\left( \mathcal{N}_{\gamma }(z\bar{z}%
)\right) ^{-\frac{1}{2}}\sum_{n=0}^{+\infty }\frac{\bar{z}^{n}}{\sqrt{%
x_{n}^{\gamma }!}}\left| \phi _{n}\right\rangle ,\ \ n=0,1,2,...\text{ },
\label{cs}
\end{equation}
where
\begin{equation}
x_{n}^{\gamma }!:=n!(n+\gamma )(2\gamma +1)(2\gamma +2)\cdots (2\gamma +n-1)
\label{4.3}
\end{equation}
and $\left\{ \left| \phi _{n}\right\rangle \right\} $ is the orthonormal
basis of an arbitrary Hilbert
space $\mathcal{H}$. From the condition
\begin{equation}
1=\left\langle z;\gamma |z;\gamma \right\rangle =2\left( \mathcal{N}_{\gamma
}(z\bar{z})\right) ^{-1}\sum_{n=0}^{+\infty }\frac{(\gamma )_{n}}{(\gamma
+1)_{n}(2\gamma )_{n}}\frac{(z\bar{z})^{n}}{n!},
\end{equation}
we see that the normalization factor is given by
\begin{equation}
\mathcal{N}_{\gamma }\left( z\bar{z}\right) =2\ _{1}F_{2}\left(
\begin{array}{c}
\gamma  \\
\gamma +1,2\gamma
\end{array}
\mid z\bar{z}\right),  \label{4.4}
\end{equation}
$_{1}F_{2}$ being the generalized hypergeometric function. It may be mentioned that the hypergeometric series $_{p}F_{q}(a_1,...,a_p;b_1,...,b_q;x)$ converge for all values of $x$ when $p\leq q$ ([11],p. 8).\\
\newline
We now give a measure with respect to which the NLCS $(\ref{cs})$ ensure the
resolution of the identity of $\mathcal{H}$ (see Appendix
A).\newline
\newline
\textbf{Proposition 3.2.} \emph{Let} $\gamma \in \lbrack 0,\infty )$.\emph{\
Then, the NLCS} \emph{$(\ref{cs})$} \emph{satisfy the following
resolution of the identity}
\begin{equation}
\int_{\mathbb{C}}\left| z;\gamma \right\rangle \left\langle z;\gamma \right|
d\mu _{\gamma }(z)=\mathbf{1}_{\mathcal{H}},  \label{identityresolution}
\end{equation}
\emph{where}
\begin{equation}
d\mu _{\gamma }(z)=\frac{4}{\Gamma (2\gamma +1)}\ _{1}F_{2}\left(
\begin{array}{c}
\gamma  \\
\gamma +1,2\gamma
\end{array}
\mid z\bar{z}\right) G_{13}^{30}\left( z\bar{z} \  \Bigg\vert \  { \gamma-1 \atop 0,\ \gamma, 2\gamma-1} \right) d\mu (z),  \label{4.7}
\end{equation}
$G_{13}^{30}(.)$ \emph{is the Meijer's G-function and $d\mu $ being the
Lebesgue measure on $\mathbb{C}$.}
\\
\\
\textbf{Remark 3.1.} When $\gamma =0$, the sequence in $(\ref
{positivenumbers})$ reduces to $x_{n}^{0}=n^{2}$ and $%
x_{n}^{0}!=(n!)^{2}$, therefore the obtained NLCS are of Barut-Girardello
type $(\ref{3.13})$ with $2\sigma =1.$ In this case, results
on overcompletness or undercompletness of discrete sets of CS
based on the use of theorems that relate the growth of analytic functions to
the density of their zeros were obtained in [12]. While at the limit $\gamma=\infty $, the
generalized factorial $(\ref{4.3})$ behaves like $(2\gamma )^{n}n!$ and one can identify (up to a scale factor) the resulting NLCS as the
canonical CS (2.1).
\newpage
\section{NLCS for the symmetric P\"{o}schl-Teller oscillator}

We recall [13] the one dimensional P\"{o}schl-Teller oscillator whose
Hamiltonian is given by
\begin{equation}
H_{\nu }=-\frac{1}{2m_{\ast }}\frac{d^{2}}{d\theta ^{2}}+V_{\nu }(\theta ),\
\
\end{equation}
the potential is 
\begin{equation}
V_{\nu }(\theta ):=\frac{\hbar ^{2}\alpha ^{2}}{2m_{\ast }}\frac{\nu (\nu -1)%
}{\cos ^{2}\alpha \theta },
\end{equation}
where $-\pi /2\alpha \leq \theta \leq \pi /2\alpha $, $\hbar $ the Planck's
constant, $\alpha >0$ is related to the range of the potential, $m_{\ast }$
is the reduced mass of the particle, $\nu >1$ is related to the potential
strength and $\theta $ gives the relative distance from the equilibrium
position. The Schr\"{o}dinger eigenvalue equation reads
\begin{equation}
H_{\nu }\phi _{n}^{\nu }=E_{n}^{\nu }\phi _{n}^{\nu },  \label{ernergy1}
\end{equation}
where the energy of a bound state is given by
\begin{equation}
E_{n}^{\nu }=\frac{\hbar^2\alpha^2}{2m_{\ast}}(\nu+n)^2.
\label{energy}
\end{equation}
Eigenfunctions corresponding to eigenvalues $(\ref{energy})$ are written as
\begin{equation}
\left\langle \theta |\phi _{n}^{\nu }\right\rangle =\sqrt{\frac{\alpha
n!(n+\nu )\Gamma (\nu )\Gamma (2\nu )}{\pi ^{1/2}\Gamma (n+2\nu )\Gamma (\nu
+1/2)}}\cos ^{\nu }(\alpha \theta )\ C_{n}^{(\nu )}(\sin \alpha \theta )
\label{22.5}
\end{equation}
in terms of Gegenbauer polynomials $C_{n}^{\nu }(.)$ and constitute an
orthogonal basis of the Hilbert space $\mathcal{H}_{\alpha }=L^{2}\left( [-%
\frac{\pi }{2\alpha },\frac{\pi }{2\alpha }],d\theta \right) $. \\ \\
\textbf{Remark 4.1.}
Observe that as $\nu \rightarrow 1$, the potential, energy levels, and
normalized eignefunctions become exactly those for the infinite square well
potential with barriers at $\theta =\pm \pi /2\alpha $. In this case, the
wave functions $\left( 4.5\right) $ become
\begin{equation}
\left\langle \theta |\phi _{n}^{1}\right\rangle =\sqrt{\frac{2\alpha }{\pi }}%
\cos (\alpha \theta )\ U_{n}(\sin \alpha \theta ),  \label{2.27}
\end{equation}
where $U_{n}(.)$ is the Chebychev polynomials written in terms of Gegenbauer
polynomials by the relation $C_{n}^{1}(x)=U_{n}(x)$, see [14].\\ \\
\textbf{Remark 4.2.}
Note also [13] that by first subtracting the zero point energy $\nu(\nu
-1)\hbar \alpha^2 /2$ and then taking limits $\nu \rightarrow \infty $, $%
\alpha \rightarrow 0$, but such that $\alpha^2\nu=m\omega/\hbar$, the potential, energy levels, and
normalized wave function become those for the harmonic oscillator.
\newpage
\textbf{Definition 4.1.} \emph{Let $\gamma \in (0,\infty )$ and $\nu >1$ be
fixed parameters. Define a set of NLCS by the following superposition}
\begin{equation}
\left| z;\gamma ,\nu \right\rangle :=\left( \mathcal{N}_{\gamma }(z\bar{z}%
)\right) ^{-\frac{1}{2}}\sum_{n=0}^{+\infty }\frac{\bar{z}^{n}}{\sqrt{%
x_{n}^{\gamma }!}}\left| \phi _{n}^{\nu }\right\rangle   \label{cs5}
\end{equation}
\emph{where $\mathcal{N}_{\gamma }(.)$ is a normalization factor}\emph{, $%
x_{n}^{\gamma }!$ is given by $(\ref{4.3})$} \emph{and} $ \left| \phi
_{n}^{\nu }\right\rangle $ \emph{are the eigenfunctions defined in}
$(\ref{22.5})$. \newline
\newline
A closed form of $(\ref{cs5})$ can be obtained in the following case
(see Appendix B).\newline
\newline
\textbf{Proposition 4.1.}\emph{\ Let $\gamma \in (0,\infty )$ and $\nu >1$.
Assuming that $\nu =\gamma $, then the wave functions of NLCS $(\ref{cs5})$}%
\emph{\ are of the form}
\begin{equation}
\left\langle \theta |z;\gamma \right\rangle =2^{\gamma -1}\sqrt{\frac{\alpha
\Gamma (\gamma +1)\Gamma \left( \gamma +\frac{1}{2}\right) }{\pi ^{1/2}}}\
\left( _{1}F_{2}\left(
\begin{array}{c}
\gamma  \\
\gamma +1,2\gamma
\end{array}
\mid z\bar{z}\right) \right) ^{-1/2}  \label{wf}
\end{equation}
\begin{equation*}
\times \bar{z}^{\frac{1}{2}-\gamma }\exp \left( {\bar{z}\sin \alpha \theta }%
\right) \ J_{\gamma -\frac{1}{2}}(\bar{z}\cos \alpha \theta )\sqrt{\cos
\alpha \theta },
\end{equation*}
for every $\theta \in \left[-\frac{\pi }{%
2\alpha },\frac{\pi }{2\alpha }\right]$. \emph{For} $\gamma =1$\emph{, which corresponds to the infinite square well
potential, wave functions are given by}
\begin{equation}
\left\langle \theta |z\right\rangle =\sqrt{\frac{\alpha }{\pi }}\left(
I_{0}(2|z|-1)\right) ^{-\frac{1}{2}}\exp \left( \bar{z}\sin \alpha \theta
\right) \sin (\bar{z}\cos \alpha \theta )  \label{5.3}
\end{equation}
\emph{in terms of the modified Bessel function of the first kind $I_{0}(.)$.}
\newline
\newline
Note that the reproducing kernel which arises from the NLCS $(\ref
{cs5})$ is
\begin{equation}
K(z,w)=\sum_{n=0}^{+\infty }\frac{(z\overline{w})^{n}}{x_{n}^{\gamma }!},
\end{equation}
and the corresponding reproducing kernel Hilbert space, denoted here by $%
\mathcal{A}_{\gamma }(\mathbb{C})$ is a subspace, consisting of functions
which are holomorphic in the domain $\mathcal{D}$, of the larger Hilbert
space $L^{2}(\mathcal{D},d\nu _{\gamma })$. Here, $\mathcal{D}=\mathbb{C}$
the whole complex plane and the measure $d\nu _{\gamma }(z,\bar{z})$ is
given by
\begin{equation}
d\nu _{\gamma }(z,\bar{z})=\frac{2}{\Gamma (2\gamma +1)}G_{13}^{30}\left( z\bar{z} \  \Bigg\vert \  { \gamma-1 \atop 0,\ \gamma, 2\gamma-1} \right) d\mu (z),
\end{equation}
moreover, it is easy to see that a non zero function $f(z)=\sum\limits_{n=0}^{+\infty }a_{n}z^{n}$
belongs to $\mathcal{A}_{\gamma }(\mathbb{C})$ if and only if the sequence $%
a_{n}$ satisfies the growth condition
\begin{equation}
\frac{1}{\Gamma (2\gamma +1)}\sum_{n=0}^{+\infty }n!(n+\gamma )\Gamma
(n+2\gamma )|a_{n}|^{2}<+\infty .
\end{equation}
In view of the resolution of the identity $(\ref{identityresolution})$, we
easily see that the map $\mathcal{B}_{\gamma }:\mathcal{H}_{\alpha
}\rightarrow \mathcal{A}_{\gamma }(\mathbb{C})$ defined by
\begin{equation}
\mathcal{B}_{\gamma }[\phi ](z)=\left( \mathcal{N}(z\bar{z})\right)
^{1/2}\langle \phi |z,\gamma \rangle _{\mathcal{H}_{\alpha}}  \label{6.5}
\end{equation}
is unitary, embedding $\mathcal{H}_{\alpha }$ into the holomorphic subspace $A_{\gamma}(\mathbb{C})\subset L^{2}\left( \mathcal{D},d\nu _{\gamma }\right) $. In order to express it as
an integral transform we make use of proposition 4.1.\newline
\newline
\textbf{Theorem 4.1.} \emph{Let} $\gamma >1$ \emph{be a fixed parameter}.
\emph{The Bargmann transform is the unitary map} $\mathcal{B}_{\gamma }:%
\mathcal{H}_{\alpha }\rightarrow \mathcal{A}_{\gamma }(\mathbb{C})$ \emph{%
defined by means of $(\ref{6.5})$ as}
\begin{equation}
\mathcal{B}_{\gamma }[\varphi ](z)=\sqrt{\alpha \frac{\Gamma (\gamma
+1)\Gamma (\gamma +1/2)}{\pi ^{1/2}}}\left( \frac{z}{2}\right) ^{\frac{1}{2}%
-\gamma }\int_{\frac{-\pi }{2\alpha }}^{\frac{\pi }{2\alpha }}\exp \left(
z\sin \alpha \theta \right) J_{\gamma -1/2}(z\sin \alpha \theta )\sqrt{\cos
\alpha \theta }\varphi (\theta )d\theta .  \label{6.6}
\end{equation}
\emph{In particular, at the limit $\gamma =1$ which corresponds to the
infinite square well potential,}
\begin{equation}
\mathcal{B}_{1}[\varphi ](z)=\frac{\left( \frac{\alpha }{\pi }\right) ^{1/2}%
}{z}\int_{\frac{-\pi }{2\alpha }}^{\frac{\pi }{2\alpha }}\exp \left( z\sin
\alpha \theta \right) \sin (z\cos \alpha \theta )\varphi \left( \theta
\right) d\theta
\end{equation}
\emph{for every} $z\in \mathbb{C}$.\newline
\newline
With the help of this transform we see that any arbitrary state $|\phi
\rangle $ in $\mathcal{H}_{\alpha }$ has a representation in terms of the NLCS $(\ref{cs5})$ as follows
\begin{equation}
|\phi \rangle =\int_{\mathbb{C}}d\mu _{\gamma }(z)\mathcal{B}_{\gamma }[\phi
](z)|z,\gamma \rangle .  \label{5.18}
\end{equation}
Therefore, the norm square of $%
|\phi \rangle $ also reads
\begin{equation}
\langle \phi |\phi \rangle _{\mathcal{H}_{\alpha }}=\frac{2}{\Gamma (2\gamma
+1)}\int_{\mathbb{C}}\left| \mathcal{B}_{\gamma }[\phi ](z)\right|
^{2}G_{13}^{30}\left( z\bar{z} \  \Bigg\vert \  { \gamma-1 \atop 0,\ \gamma, 2\gamma-1} \right) \sqrt{\mathcal{N}_{\gamma }(z\bar{z})}d\mu (z).
\end{equation}
\newline
\textbf{Remark 4.3.} An expression generalizing the above coefficients $%
x_{n}^{\gamma }$ in (4.1) have been considered in [15] where the authors
have provided an algebraic construction of the coherent states for a wide
class of potentials, belonging to the confluent hypergeometric and
hypergeometric classes. \\ \\
\textbf{Remark 4.4.} Note also that in a similar context $\left[16%
\right] $ the authors were dealing with coherent states for the Hamiltonian
with the P\"{o}schl-Teller potential, for which they were investigating nonclassical properties through statistics of the
corresponding photon-counting probability distribution.
\newpage
\section{Orthogonal polynomials attached to NLCS}

Following [2], there are two sets of orthogonal polynomials, we can
associate with the family of NLCS $(\ref{cs})$ in the
following way.

\subsection{Polynomials attached to the measure $d\protect\nu _{\protect%
\gamma }$}

These polynomials are obtained by symmetrizing the measure
\begin{equation}
d\nu _{\gamma }(r)=\frac{2}{\Gamma (2\gamma +1)}G_{13}^{30}\left( r^2 \  \Bigg\vert \  { \gamma-1 \atop 0,\ \gamma, 2\gamma-1} \right) rdr,  \label{4.11}
\end{equation}
in $(\ref{4.7})$ giving the identity
\begin{equation}
d\eta _{\gamma }(t)=\frac{1}{2}d\nu _{\gamma }(|t|)
\label{symmetrisemeasure}
\end{equation}
on the symmetric interval $(-\infty ,+\infty )$, with moments
\begin{equation}
\mu _{2n}=2\int_{0}^{\infty }r^{2n}d\eta _{\gamma }(t)=x_{n}^{\gamma }!,\ \
\mu _{2n+1}=2\int_{0}^{\infty }r^{2n+1}d\eta _{\gamma }(t)=0,\quad
n=0,1,2,...\text{ .}  \label{momentproblem}
\end{equation}
Precisely, a set of (monic) polynomials $P_{n}(t)$, $n=0,1,2,...,$
orthogonal with respect to the measure $d\eta _{\gamma }$, are defined using
the Hankel determinant

\begin{equation}
P_{n}(x)=\frac{1}{\Delta _{n-1}}
\begin{vmatrix}
\mu _{0} & \mu _{1} & \cdots  & \mu _{n} \\
\vdots  & \vdots  & . & \vdots  \\
\mu _{n-1} & \mu _{n} & \cdots  & \mu _{2n-1} \\
1 & x & \cdots  & x^{n}
\end{vmatrix}
,\ \ \ \Delta _{n}=
\begin{vmatrix}
\mu _{0} & \mu _{1} & \cdots  & \mu _{n} \\
\mu _{1} & \mu _{2} & \cdots  & \mu _{n+1} \\
\vdots  & \vdots  & . & \vdots  \\
\mu _{n} & \mu _{n+1} & \cdots  & \mu _{2n}
\end{vmatrix}
.  \label{matrix}
\end{equation}
We will discuss the particular values $\gamma =0$ and $\gamma =+\infty .$
The case $\gamma =1$ is of a particular interest because it is associated
with NLCS for the infinite square well potential.\newline
\newline
\textbf{Case} $\gamma =0.$ As mentioned above the obtained NLCS are of
Barut-Girardello with $x_{n}!=(n!)^{2}$. In this case, the measure in $(\ref
{4.11})$ takes the form $d\nu _{0}(r)=4K_{0}(2r)rdr$, by using the relation
([17], p.61):
\begin{equation}
G_{02}^{20}\left( y|\alpha ,\beta \right) =2y^{\frac{\alpha +\beta }{2}%
}K_{\alpha -\beta }(2\sqrt{y}),  \label{meijermeijer}
\end{equation}
for $y=r^{2}$ and $\alpha =\beta =0$, where
\begin{equation}
K_{0}(\rho )=\int_{0}^{+\infty }\frac{\cos (\rho t)}{\sqrt{t^{2}+1}}dt,\ \ \
\rho >0,
\end{equation}
is the MacDonald function of order zero ([9], p.183). Then, the measure
$(\ref{symmetrisemeasure})$ reads
\begin{equation}
d\eta _{0}(t)=2K_{0}(2|t|)|t|dt,\quad t\in (-\infty ,+\infty ),
\label{symmmetrieingmeasure}
\end{equation}
and the moment problem $(\ref{momentproblem})$ takes the form
\begin{equation}
\mu _{2n}=2\int_{0}^{+\infty }t^{2n}d\eta _{0}(t)=(n!)^{2},\ \ \mu
_{2n+1}=2\int_{0}^{\infty }t^{2n+1}d\eta _{0}(t)=0,\quad n=0,1,2,...\text{ .}
\label{momen}
\end{equation}
As illustration, the first polynomials are given by
\begin{eqnarray*}
P_{0}(x) &=&1 \\
P_{1}(x) &=&x \\
P_{2}(x) &=&x^{2}-1 \\
P_{3}(x) &=&x^{3}-4x \\
P_{4}(x) &=&x^{4}-\frac{32}{3}x^{2}+\frac{20}{3} \\
P_{5}(x) &=&x^{5}-\frac{108}{5}x^{3}+\frac{252}{5}x \\
P_{6}(x) &=&x^{6}-\frac{1593}{41}x^{4}+\frac{9612}{41}x^{2}-\frac{4716}{41}.
\end{eqnarray*}
\begin{figure}[H]
\centering
\includegraphics[width=0.65\textwidth]{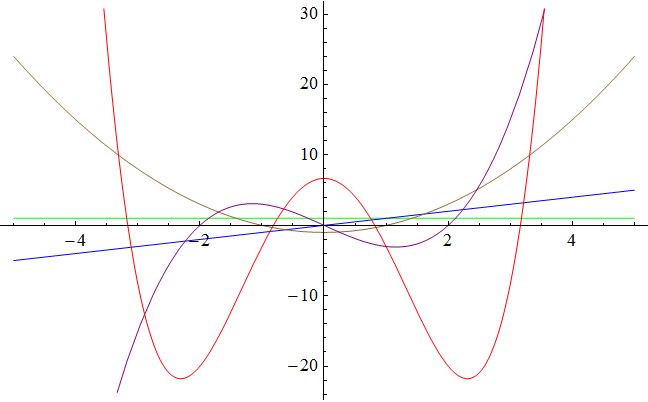}
\caption{{The polynomials $P_0,P_1,P_2,P_3$ and $P_4$}}
\end{figure}
These polynomials are symmetric with respect to the origin and satisfy the
orthogonality relations
\begin{equation}
\int_{-\infty }^{+\infty }P_{n}(x)P_{m}(x)d\eta _{0}(x)=\xi _{n}\delta _{mn},
\label{orthooo}
\end{equation}
where $\xi _{n}>0$ is a normalization constant and $\delta _{mn}$ is the
Kronecher's symbol. The normalized polynomials
\begin{equation}
\widetilde{P}_{n}(x):=\frac{1}{\sqrt{\xi _{n}}}P_{n}(x),  \label{polypntilde}
\end{equation}
satisfy a three-terms recurrence relation ([18], p.240):
\begin{equation}
x\widetilde{P}_{n}(x)=A_{n+1}\widetilde{P}_{n+1}(x)+A_{n}\widetilde{P}%
_{n-1}(x)
\end{equation}
where the coefficient $A_{n}$ obey the asymptotic formula
\begin{equation}
\lim_{n\rightarrow \infty }\frac{A_{n}}{n}=\frac{\pi }{16}
\end{equation}
which constitutes the property on the $A_{n}$'s we know up to now [18]. In fact,
by taking
\begin{equation}
V_{n}(x):=\widetilde{P}_{2n}(x),  \label{polyvandvtilde}
\end{equation}
Eq.$(\ref{orthooo})$ gives the orthogonality relations
\begin{equation}
2\int_{0}^{\infty }V_{n}(x)V_{m}(x)K_{0}(2\sqrt{x})dx=\delta _{mn}.
\end{equation}
Straightforward calculations using the moments formula
\begin{equation}
2\int_{0}^{+\infty }K_{0}(2\sqrt{x})x^{n}dx=(n!)^{2},  \label{momentfinal}
\end{equation}
provide us with the exact constants
\begin{equation}
\xi _{2}=3,\ \xi _{4}=656/3\ \text{and}\ \xi _{6}=3681936/41,
\end{equation}
corresponding respectively to polynomials $V_{2},\ V_{4}$ and $V_{6}$. So
that we recover the first three polynomials as given by Ditkin and Prudnikov
([18], p.240):
\begin{eqnarray*}
V_{1}(x) &=&\frac{x-1}{\sqrt{3}} \\
V_{2}(x) &=&\sqrt{\frac{3}{41}}\left( \frac{1}{4}x^{2}-\frac{8}{3}x+\frac{5}{%
3}\right)  \\
V_{3}(x) &=&\sqrt{\frac{41}{2841}}\left( \frac{1}{36}x^{3}-\frac{177}{164}%
x^{2}+\frac{267}{41}x-\frac{131}{41}\right) .
\end{eqnarray*}
\ \newline
\textbf{Case} $\gamma =1.$ We now proceed to attach a set of orthogonal
polynomials, say $Q_n(x)$, to NLCS for the infinite square well potential. The
corresponding generalized factorial takes the form $x_{n}^{1}!=((n+1)!)^{2},%
\ n=0,1,2,...$ $.$ The measure in $(\ref{4.11})$ can be written as $d\nu
_{1}(r)=G_{02}^{20}\left( r^{2}|1,1\right) rdr$. By the help of  $(%
\ref{meijermeijer})$ the measure in $(\ref{symmetrisemeasure})$ reads $d\eta
_{1}(t)=\frac{1}{2}d\nu _{1}(|t|)$, $t\in (-\infty ,+\infty ).$ Therefore, the
moment problem $(\ref{momentproblem})$ takes the form
\begin{equation}
\mu _{2n}=\int_{0}^{+\infty }t^{n+\frac{1}{2}}K_{0}\left( 2\sqrt{t}\right)
dt=((n+1)!)^{2},\ \mu _{2n+1}=\int_{0}^{+\infty }t^{n+1}K_{0}\left( 2\sqrt{t}%
\right) dt=0\text{.}
\end{equation}
The polynomials $Q_{n}(x),\ n=0,1,2,...$ $,$ orthogonal with respect to the
measure $d\eta _{1}$ can be computed using $(\ref{matrix})$. The first
polynomials are given by{\small {
\begin{eqnarray*}
Q_{0}(x) &=&1 \\
Q_{1}(x) &=&x \\
Q_{2}(x) &=&x^{2}-4 \\
Q_{3}(x) &=&x^{3}-9x \\
Q_{4}(x) &=&x^{4}-\frac{108}{5}x^{2}+\frac{252}{5} \\
Q_{5}(x) &=&x^{5}-\frac{256}{7}x^{3}+\frac{1296}{7}x \\
Q_{6}(x) &=&x^{6}-\frac{8208}{131}x^{4}+\frac{37429}{50}x^{2}-\frac{21035}{16%
}.
\end{eqnarray*}}
\begin{figure}[H]
$${\small { \includegraphics[width=0.65%
\textwidth]{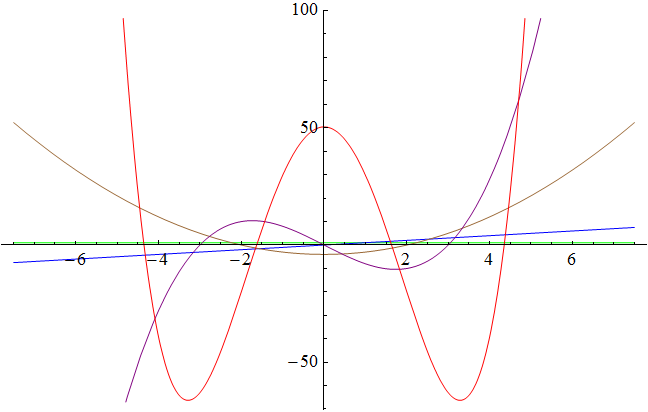} }  }$$
\caption{{The polynomials $Q_0,Q_1,Q_2,Q_3$ and $Q_4$}}
\end{figure}
}We also note that polynomials $Q_{n}(x)$ with even degree can be connected
with polynomials $P_{n}(x)$ in $(\ref{polypntilde})$ with odd degree by $%
xQ_{2n}(x)=P_{2n+1}(x)$, $n=0,1,2,...$ $.$\newline
\newline
\textbf{Case} $\gamma =+\infty $. This case corresponds (up to a scale
factor) to the canonical CS with the measure $(\ref
{canonicalcoherent})$. The resulting orthogonal polynomials are
found to be the Laguerre polynomials ($\left[ 2\right] $, p.5).\newline \\
\textbf{Remark 5.2}. For different values of $\gamma \in $ $[0,\infty )$
these polynomials can be associated with the Ditkin-Prudnikov problem ([18],
pp.239-240) as follows. Let $V_{0}(x,k)=1$, $V_{1}(x,k),...,V_{n}(x,k)$, $k$
a positive integer, be the orthogonal system of polynomials on the interval $%
0\leq x\leq \infty $, with respect to the\textit{\ ultra-exponential} weight
function
\begin{equation}
\xi (x,k)=\frac{1}{2\pi i}\int_{a-i\infty }^{a+i\infty }x^{-s}\Gamma
^{k}(s),\ \ a,\ x,\ \Re s>0.
\end{equation}
That is,\
\begin{equation}
\int_{0}^{\infty }V_{n}(x,k)V_{m}(x,k)\xi (x,k)dx=\delta _{nm}.
\end{equation}
We note that building the generating function, an analogue of Rodrigues
formula, the recurrence relation for orthogonal polynomials $V_{n}(x,k)$, $%
k\geq 2$, is still an open problem ([18], pp.239-240). Now let us set our
parameter $\gamma =\frac{2-k}{k-1}$. When $k=1\ (\gamma =\infty ),$ then $%
\xi (x,1)=e^{-x}$ and $V_{n}(x,1)=(-1)^{n}L_{n}(x)$ are Laguerre
polynomials. If $k=2\ (\gamma =0)$, then $\xi (x,2)=2K_{0}(2\sqrt{x})$ and
polynomials $V_{n}(x,2)$ are those connected to the $P_{n}(x)$ in $(\ref
{polyvandvtilde})$. Now, since the case $\gamma =1$ which was involved in
the infinite square well potential (see $(\ref{2.27})$ above) corresponds to
a fractional value $k=3/2$, it may be useful to extend the Ditkin-Prudnikov
problem to values $k\in ]1,2[$.{\small \ }

\subsection{Polynomials associated to shift operators}

A second set of polynomials can be associated with the sequence $x_{n}$
defining the NLCS as follows. Define the formal shift operator
\begin{equation}
a\phi _{n}=\sqrt{x_{n}}\phi _{n-1},\ \ \ a\phi _{0}=0,\ \ \ a^{\ast }\phi
_{n}=\sqrt{x_{n+1}}\phi _{n+1},\ \ \ n=0,1,2,...\text{ }.
\end{equation}
Then, if $\sum_{n=0}^{\infty }\frac{1}{\sqrt{x_{n}}}=\infty $, the operator $%
Q=\frac{1}{\sqrt{2}}\left( a+a^{\ast }\right) $ is essentially self-adjoint
and hence has a unique self-adjoint extension [19-20] which we again
denote by $Q$. This operator acts on the basis vector $\phi _{n}$ as
follows\
\begin{equation}
Q\phi _{n}=\sqrt{\frac{x_{n}}{2}}\phi _{n-1}+\sqrt{\frac{x_{n+1}}{2}}\phi
_{n+1}.  \label{5.21}
\end{equation}
There exists an even measure $dw$ such that $Q$ acts on the space $%
L^{2}(\mathbb{R},dw)$ as the operator of multiplication and the $\phi _{n}$ are
functions in this space in which Eq.$(\ref{5.21})$ reads
\begin{equation}
x\phi _{n}(x)=\sqrt{\frac{x_{n}}{2}}\phi _{n-1}(x)+\sqrt{\frac{x_{n+1}}{2}}%
\phi _{n+1}(x),\quad n=1,2,...\,,  \label{recurrencerelationphi}
\end{equation}
with initial conditions, $\phi _{-1}=0$ and$\ \ \phi _{0}= 1$. The measure $dw$
comes from the spectral family of projectors, $E_{x},\ x\in \mathbb{R}$, of the
operator $Q$, in the sense that $dw(x)=\left\langle \phi _{0}|E_{x}\phi
_{0}\right\rangle $.\newline
{\small \newline
}\textbf{Case} $\gamma =0.$ This case corresponds to the sequence $%
x_{n}=n^{2}$ and to CS of Barut-Girardello type (with $\sigma
=1/2$ in $\left( 2.10\right) $). Here, the associated polynomials, say $\phi
_{n}^{(1/2)},$ satisfy the recurrence relation
\begin{equation}
x\phi _{n}^{(1/2)}(x)=\frac{n+1}{\sqrt{2}}\phi _{n+1}^{(1/2)}(x)+\frac{n}{%
\sqrt{2}}\phi _{n-1}^{(1/2)}(x).  \label{5.25}
\end{equation}
By $(\ref{5.25})$ we can compute them successively. Here, we give the first polynomials
\begin{eqnarray*}
\phi _{0}^{(1/2 )}(x)&=& 1\\
\phi _{1}^{(1/2 )}(x) &=& 2 x\\
\phi _{2}^{(1/2 )}(x) &=& 2 x^2 - 1\\
\phi _{3}^{(1/2 )}(x)&=& 4 x^3 -\frac{8}{3} x\\
\phi _{4}^{(1/2 )}(x)&=& 2 x^4 - \frac{10}{3} x^2 + 1\\
\phi _{5}^{(1/2 )}(x) &=& \frac{4}{5} x^5 - \frac{16}{5} x^3 + \frac{46}{15} x\\
\phi _{6}^{(1/2 )}(x) &=& \frac{4}{15} x^6 - \frac{56}{15} x^4 + \frac{196}{45} x^2 - 1
\end{eqnarray*}
\begin{figure}[H]
\small {\includegraphics[width=0.65%
\textwidth]{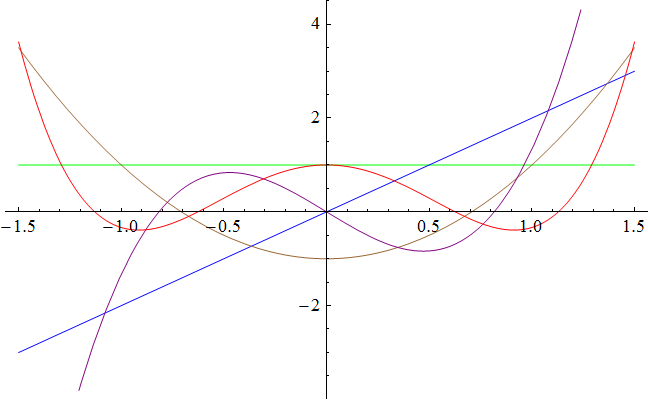} }
\caption{{The polynomials $\phi_0,\phi_1,\phi_2,\phi_3$ and $\phi_4$}}
\end{figure}
To obtain more information on these polynomials, we establish the following
result. \newline
{\small \newline
}\textbf{Proposition 5.1.} \textit{The polynomials satisfying} $(\ref{5.25})$\textit{\ are a special case of
Meixner-Pollaczeck polynomials}{\small \
\begin{equation}
\phi _{n}^{\left( 1/2\right) }(x)=P_{n}^{(1/2)}\left( \frac{x}{\sqrt{2}},%
\frac{\pi }{2}\right)   \label{Phi}
\end{equation}
}\textit{where}{\small \emph{\ }
\begin{equation}
P_{n}^{(1/2)}(u,\pi /2)=i^{n}{}_{2}F_{1}\left(
\begin{array}{c}
-n,\frac{1}{2}+iu \\
1
\end{array}
\mid 2\right)
\end{equation}
}\textit{is given in terms of a terminating Gauss hypergeometric }$_{2}F_{1}$%
\textit{-sum. \newline
}{\small \newline
}\textbf{Proof.}{\small \textbf{\ }}We consider the normalized polynomials
\begin{equation}
q_{n}(x):=\frac{n!}{2^{\frac{1}{2}n}}\phi _{n}^{(1/2)}(x)  \label{tocto}
\end{equation}
which satisfy the recurrence relation\
\begin{equation}
q_{n+1}(x)-xq_{n}(x)+\frac{1}{2}n^{2}q_{n-1}(x)=0.  \label{recursio}
\end{equation}
Multiplying\ $(\ref{recursio})$\ by $t^{n}/n!$ and summing over $n$, we
obtain that\textit{\ }
\begin{equation}
\sum_{n=0}^{+\infty }q_{n+1}(x)\frac{t^{n}}{n!}-x\sum_{n=0}^{+\infty
}q_{n}(x)\frac{t^{n}}{n!}+\frac{1}{2}\sum_{n=0}^{+\infty }nq_{n-1}(x)\frac{%
t^{n}}{(n-1)!}=0.  \label{rduct}
\end{equation}
Setting\
\begin{equation}
G_{x}(t):=\sum_{n=0}^{+\infty }q_{n}(x)\frac{t^{n}}{n!},  \label{condition}
\end{equation}
then $(\ref{rduct})$ leads to the differential equation
\begin{equation}
(t^{2}+2)\frac{d}{dt} G_{x}(t)+(t-2x)G_{x}(t)=0,
\end{equation}
which, by using the condition $G(0,0)=1$, gives that\
\begin{equation}
\label{222.1}
G_{x}(t)=\frac{\sqrt{2}}{\sqrt{2+t^{2}}}\exp \left( \sqrt{2}x\arctan \frac{t%
}{\sqrt{2}}\right) .
\end{equation}
By another hand, if we particularize the
generating function of the Meixner-Pollaczeck polynomials ([21], p.8):
\begin{equation}
\sum_{n=0}^{+\infty }P_{n}^{(\lambda )}(u,\phi )t^{n}=\left( 1-e^{i\phi
}t\right) ^{-\lambda +iu}\left( 1-e^{-i\phi }t\right) ^{-\lambda -iu}.
\label{poll-generating}
\end{equation}
by setting $\lambda =1/2$,$\ \ \phi =\pi /2$ and $u=x/\sqrt{2}$, we then
obtain
\begin{equation}
\sum_{n=0}^{+\infty }P_{n}^{(1/2)}\left( \frac{x}{\sqrt{2}},\frac{\pi }{2}%
\right) t^{n}=\frac{1}{\sqrt{1+t^2}}\left( \frac{1-it}{1+it}\right) ^{\frac{i}{\sqrt{2}}x}.
\end{equation}
Next, using the identity\
\begin{equation}
\left( \frac{1-it}{1+it}\right) ^{\frac{1}{2}iz}=\exp \left( z\arctan
t\right) ,
\label{ident}
\end{equation}
for $z=\sqrt{2}x$, we get that
\begin{equation}
\sum_{n=0}^{+\infty }P_{n}^{(1/2)}\left( \frac{x}{\sqrt{2}},\frac{\pi }{2}%
\right) t^{n}=\frac{1}{\sqrt{1+t^{2}}}\exp\left(\sqrt{2} x \arctan t\right).
\label{Meixner}
\end{equation}
\ \\
By comparing $(\ref{Meixner})$ with $(\ref{222.1})$, we arrive at $(\ref
{Phi})$. This completes the proof. $\Box $\\
\newpage
\textbf{Remark 5.3.} In the case of the sequence $%
x_{n}=n\left( n+2\sigma -1\right) $ (which coincides with our sequence $%
x_{n}^{0}=n^2$ when $2\sigma =1$) with $2\sigma=2,3... $, we can use similar calculations to show that
the resulting polynomials, say $\phi _{n}^{\left( \sigma \right) }(x),$ have
the generating function
\begin{equation*}
\sum\limits_{n\geq 0}t^{n}\phi _{n}^{(\sigma )}(x)=\left( 1+t^{2}\right)
^{-\sigma }\exp \left( \sqrt{2}x\arctan t\right)
\end{equation*}
and therefore, they are the Meixner-Pollaczeck polynomials
\begin{equation*}
\phi _{n}^{(\sigma )}(x)=P_{n}^{\left( \sigma \right) }\left( \frac{x}{\sqrt{%
2}},\frac{\pi }{2}\right) .
\end{equation*}
Note that these polynomials occur in the expression of eigenstates of the
relativistic linear oscillator $\left[22\right] $.\\
\\
\textbf{Case} $\gamma =1.$ This corresponds to the sequence $x_{n}=(n+1)^{2}$
which is related to the infinite square well potential. Here the attached polynomials $\phi _{n}$ satisfy the three-terms recurrence relation
\begin{equation}
x\phi _{n}(x)=\frac{n+2}{\sqrt{2}}\phi _{n+1}(x)+\frac{n+1}{\sqrt{2}}\phi
_{n-1}(x).  \label{gamma1}
\end{equation}
\textbf{Proposition 5.2.} \emph{The polynomials $\phi_n$, denoted here by} $\phi_n^{(\frac{1}{2},1)},$ \emph{are the generalized Meixner-Pollaczeck polynomials given by }
\begin{equation}
\phi_n^{(\frac{1}{2},1)}(x)=P^{(1/2)}_n\left(\frac{x}{\sqrt{2}}, \frac{\pi}{2}, 1\right)
\end{equation}
\emph{which are orthogonal on }$(-\infty,+\infty)$ \emph{with respect to the weight function}
\begin{equation}
\label{weightfun}
\omega(x)=(2\pi)^{-1}\left|\Gamma\left(\frac{3}{2}+i\frac{x}{\sqrt{2}}\right)\right|^2\left|\ _{2}F_1\left(\frac{1}{2}+i\frac{x}{\sqrt{2}},1;\frac{3}{2}+i\frac{x}{\sqrt{2}};-1\right)\right|^{-2}
\end{equation}
\ \\
\textbf{Proof.} The result is deduced from the three-terms recurrence relation ([23], p.2256):
\begin{equation}
(n+c+1)P_{n+1}^{\lambda}(x)-2\left[(n+\lambda+c)\cos\phi+x\sin\phi\right]P_{n}^{\lambda}(x)+(n+2\lambda+c-1)P_{n-1}^{\lambda}(x)=0,
\end{equation}
where $P_n^{(\lambda)}(x):=P_n^{\lambda}(x,\phi,c)$ with $P^{(\lambda)}_{-1}(x)=0,\ P^{(\lambda)}_{0}(x)=1,\ 0<\phi<\pi,\ 2\lambda+c>0,\ c\geq0,$ or $0<\phi<\pi,\ 2\lambda+c\geq1,\ c>-1$, in the case of parameters $\lambda=1/2$ and $\phi=\pi/2$. The weight function $(\ref{weightfun})$ is obtained from ([23], p.2256) with similar replacements of parameters. $\Box$
\\ \\
\textbf{Case} $\gamma\to\infty.$ To this limit correspond the
canonical CS as mentioned above and the $\phi
_{n}(x)$ are the well-known Hermite polynomials ([2], p.6)
which appear in solutions of the Schr\"{o}dinger equation for the
harmonic oscillator.
\section{Concluding remarks}

We have replaced the factorial $n!$ occurring in coefficients $z^{n}/\sqrt{n!}$
of the canonical coherent states by a specific generalized factorial $%
x_{n}^{\gamma }!=x_{0}^{\gamma }x_{1}^{\gamma }\cdots x_{n}^{\gamma }$, where $%
x_{n}^{\gamma }$ is a sequence of positive numbers  and $\gamma \in \left(
0,\infty \right) $ being a parameter. The new coefficients are then used to
consider a superposition of eigenstates of the Hamiltonian with a symmetric P%
\"{o}schl-Teller potential depending on a parameter $\nu >1$. The
obtained states constitute a one-parameter family of nonlinear coherent
states (NLCS). For equal parameters $\gamma =\nu $, we define the associated
Bargmann-type transform and we derive some results on the infinite square
well potential. Next, we have proceeded by a general method $\left[ 2\right] $ to
discuss, for some different values of $\gamma ,$ two sets of orthogonal
polynomials that are naturally associated with these NLCS. One set of these
polynomials, say $P_{n} $, is obtained from a symmetrization
of the measure which gives the resolution of the identity for the NLCS.
Here, we can suggest a new generalization of these NLCS themselves by
replacing the coefficients $z^{n}/\sqrt{x_{n}^{\gamma }!}$ by the constructed
polynomials $P_{n}$. In this direction, it's crucial to know
some basic properties of these polynomials. However, for many values of $\gamma ,$ such properties are not known. As example, for $\gamma
=0$, the NLCS are of Barut-Girardello type and the resulting polynomials are
related to the Ditkin-Prudnikov problem which is still open$.$ \ The second
set of orthogonal polynomials, say $\phi _{n} $, arises from
the shift operators associated to these coherent states. In this case, to
polynomials $\phi _{n}$ a Hamiltonian system could be
associated $\left[ 24\right] $. Here, the ideal would be to recover the whole
structure of the NLCS from the sequence of positive numbers $x_{n}^{\gamma }$
as a unique data. However, except having the three-terms recurrence
relation, getting more informations on the $\phi _{n}$ is
not so easy. Indeed, while dealing with an example cited in $\left[ 2\right]
$ the authors $\left[25\right] $ have obtained a uniform asymptotic
expansion of $\phi _{n}$ as $n$ tends to infinity and they
have concluded that the weight function
associated with the $\phi_n$ has an usual singularity which has never
appeared for orthogonal polynomials in the Askey scheme.
\begin{appendix}
\section{The proof of proposition 3.2.}
\textbf{Proof.} Let us assume that the measure takes the form $d\mu _{\gamma}(z)=\mathcal{N}_{\gamma}(z\bar{z})h(z\bar{z})d\mu(z),$
where $h$ is an auxiliary density function to be determined. In terms of
polar coordinates $z=\rho e^{i\theta} ,\ \rho>0$ and $\theta\in [0,2\pi)$, then the measure can be rewritten as
\begin{equation}
\label{Mesure}
d\mu _{\gamma}(z)=\mathcal{N}_{\gamma}(\rho
^{2})h(\rho ^{2})\rho d\rho \frac{d\theta }{2\pi }.
\end{equation}
Using the expression $(\ref{cs})$ of coherent states, the operator
\begin{equation}
\mathcal{O}_{\gamma}=\int_{\mathbb{C}}\left| z; \gamma
\right\rangle \left\langle\gamma;z\right| d\mu _{\gamma}(z)\
\end{equation}
reads successively,
\begin{eqnarray}
\mathcal{O}_{\gamma}&=&\sum\limits_{n,m=0}^{+\infty} \left(  \int_0^{+\infty} \frac{\rho^{n+m}h(\rho^2)\rho d\rho}{\sqrt{\sigma_{\gamma}(n)\sigma_{\gamma}(m)}}\left( \int_0^{2\pi} e^{i(n-m)\theta}\frac{d\theta}{2\pi} \right)\right)  \vert \phi_n\rangle \langle \phi_m\vert\\
&=&\sum\limits_{n=0}^{+\infty}\frac{1}{n!(n+\gamma)(2\gamma+1)_{n-1}}\left(  \int_0^{+\infty}\rho^{2n}h(\rho^2) \rho d\rho \right)  \vert \phi_n\rangle \langle \phi_n\vert
\end{eqnarray}
By a change of variable, we get
\begin{eqnarray}
\label{4.14}
\mathcal{O}_{\gamma}&=&\sum\limits_{n=0}^{+\infty}\frac{1}{2n!(n+\gamma)(2\gamma+1)_{n-1}}\left(  \int_0^{+\infty}r^{n}h(r)dr \right)  \vert \phi_n\rangle \langle \phi_n\vert.
\end{eqnarray}
Now, we need to determinate the function $h$ such that
\begin{equation}
\label{Meijer}
\int_0^{+\infty}r^{n}h(r)dr=2n!(n+\gamma)(2\gamma+1)_{n-1}.
\end{equation}
For this, we recall the integral formula ([17], p.67):
\begin{equation}
\label{Meijerfirstintegral}
\int_{0}^{+\infty}G^{m l}_{p q} \left(\omega t \  \Bigg\vert \  {a_1,\cdots,a_p\atop b_1,\cdots,b_q} \right)t^{s-1}dt=\frac{1}{\omega^s}\frac{\prod\limits_{j=1}^{m}\Gamma(b_j+s)\prod\limits_{j=1}^{l}\Gamma(1-a_j-s)}{\prod\limits_{j=m+1}^{q}\Gamma(1-b_j-s)\prod\limits_{j=l+1}^{p}\Gamma(a_j+s)}
\end{equation}
involving the Meijer's function $G_{pq}^{ml}$ with conditions $0\leq l\leq p<q$; $0\leq m\leq q$; $\omega\neq0$; $c^{\ast}=m+l-\frac{p}{2}-\frac{q}{2}>0$, $|\arg\omega|<c^{\ast}\pi$; $-\min\Re(b_j)<\Re(s)<1-\max\Re(a_k)$ for $j=1,\cdots, m$ and $k=1,\cdots, l$.
For parameters $\omega=1, \ p=1, \ q=3, \ m=3, \ l=0$, $a_1=\gamma,\ b_1=1,\ b_2=\gamma+1,\ b_3=2\gamma$ and $s=n$. Equation $(\ref{Meijerfirstintegral})$ reduces to
\begin{equation}
\label{Meijerfreeintegral}
\int_{0}^{+\infty}G^{3 0}_{1 3} \left(r \  \Bigg\vert \  {\gamma\atop 1,\gamma+1, 2\gamma} \right)\frac{2r^{n-1}}{\Gamma(2\gamma+1)}dr=2n!(n+\gamma)(2\gamma+1)_{n-1}.
\end{equation}
This suggests us to take the weight function
\begin{equation}
\label{weihtfunction1}
h(r)=\frac{2r^{-1}}{\Gamma(2\gamma+1)}G^{30}_{13} \left(r \  \Bigg\vert \  {\gamma\atop 1,\gamma+1, 2\gamma} \right).
\end{equation}
By using the multiplication formula ([26], p.46):
\begin{equation}
y^{\sigma}G^{m l}_{p q} \left(y \  \Bigg\vert \  {(a_p)\atop (b_q)} \right)=G^{m l}_{p q} \left(y \  \Bigg\vert \  {(a_p+\sigma)\atop (b_q+\sigma)} \right),
\end{equation}
Eq.$(\ref{weihtfunction1})$ becomes
\begin{equation}
\label{weihtfunction2}
h(r)=\frac{2}{\Gamma(2\gamma+1)}G^{3 0}_{1 3} \left(r \  \Bigg\vert \  {\gamma-1 \atop 0,\gamma,2\gamma-1} \right).
\end{equation}
Replacing $(\ref{weihtfunction2})$ into $(\ref{Mesure})$ we arrive at the measure stated $(\ref{4.7})$. With this measure equation $(\ref{4.14})$ reduces to
\begin{equation}
\mathcal{O}_{\gamma}=\sum_{n=0}^{+\infty}\left|\phi_{n}\right\rangle\left\langle \phi_n\right|=\textbf{1}_{\mathcal{H}}.
\end{equation}
since $\{\left|\phi_{n}\right\rangle\}$ is an orthonormal basis of $\mathcal{H}$. In other words we arrive at $(\ref{identityresolution})$.
This completes the proof. $\Box$
\section{Proof of proposition 4.1.}
\textbf{Proof.} We start from $(\ref{cs5})$ by writting the expression of the wavefunction
\begin{equation}
\left\langle \theta|z;\gamma\right\rangle:=\left\langle \theta|z;\gamma,\gamma\right\rangle=\left( \mathcal{N}%
_{\gamma}\left(z\bar{z}\right) \right) ^{-\frac{1}{2}%
}\sum\limits_{n=0}^{+\infty }\frac{\bar{z}^n}{%
\sqrt{x^{\gamma}_n!}}\left\langle
\theta|\phi^{\gamma}_{n}\right\rangle.
\end{equation}
To get a closed form of the series
\begin{equation}
\mathcal{S}(\theta)=\sum_{n=0}^{+\infty}\frac{\bar{z}^n}{\sqrt{x^{\gamma}_n!}}\left\langle \theta|\phi^{\gamma}_{n}\right\rangle
\end{equation}
we replace $\left\langle \theta|\phi^{\gamma}_{n}\right\rangle$ by its expression in $(\ref{22.5})$, then we have
\begin{equation}
\mathcal{S}(\theta)=\sqrt{\frac{\alpha\Gamma(\gamma+1)}{\pi^{1/2}\Gamma\left(\gamma+\frac{1}{2}\right)}}\cos^{\gamma}(\alpha \theta)\sum_{n=0}^{+\infty}\frac{\bar{z}^n}{(2\gamma)_n}C_n^{\gamma}(\cos \alpha \theta).
\end{equation}
We now make use of the generating formula for Gegenbauer polynomials ([27], 711):
\begin{equation}
\sum_{k=0}^{+\infty}\frac{t^k}{(2\tau)_k}C_k^{\tau}(y)=\Gamma\left(\tau+\frac{1}{2}\right)e^{yt}\left(\frac{t}{2}\sqrt{1-y^2}\right)^{\frac{1}{2}-\tau}J_{\tau-\frac{1}{2}}\left(t\sqrt{1-y^2}\right)
\end{equation}
here $J_{\tau}(.)$ denotes the Bessel function of order $\tau$. For parameters $k=n,\ t=\bar{z}, \tau=\gamma $ and $y=\sin \alpha \theta$, this gives
\begin{equation}
\mathcal{S}(\theta)=2^{\gamma-1/2}\sqrt{\frac{\alpha\Gamma(\gamma+1)\Gamma\left(\gamma+\frac{1}{2}\right)}{\pi^{1/2}}}\bar{z}^{1/2-\gamma}\exp\left(\bar{z}\sin \alpha \theta\right) \ J_{\gamma-\frac{1}{2}}(\bar{z}\cos \alpha \theta)\sqrt{\cos \alpha \theta}
\end{equation}
which gives the expression $(\ref{wf})$. As mentioned above, when $\nu=1$, the symmetric PT
potential becomes the infinite square well potential with eigenfunctions $\{\phi_{n}^{1}(\theta)\}$. So that the result $(\ref{5.3})$ is deduced by setting $\gamma=1$ in the expression $(\ref{wf})$ and by using the fact $\Gamma(3/2)=\sqrt{\pi}/2$ together with the identity ([28], p.600):
\begin{equation}
\ _{1}F_2(1,2,2;\zeta^2)=\frac{1}{\zeta^2}(I_{0}(2\zeta)-1)
\end{equation}
for $\zeta=|z|$ and by using formula ([29], p.203):
\begin{equation}
J_{1/2}(\xi)=\sqrt{\frac{2}{\pi\xi}}\sin \xi,
\end{equation}
where we have chosen the variable $\xi=\bar{z}\cos \alpha\theta$. This ends the proof. $\Box$
\end{appendix}

\section*{{\protect\small {\protect\footnotesize References}}}

{\tiny {

$[1]$ V. V. Dodonov, Nonclassical states in quantum optics: a squeezed review of the first 75 years, \emph{J. Opt. B: Quantum Semiclass. Opt.} \textbf{4} R1-R33 (2002) \\
$[2]$ S. T. Ali, M. E. H. Ismail, Some orthogonal polynomials arising from
coherent states, \emph{J.Phys A: Math.} Theor, \textbf{45} (2012)\newline
$[3]$ V. V. Borzov, Orthogonal polynomials and generalized oscillator
algebras, \emph{Integral Transforms Spec. Funct.} \textbf{12} (2001)\newline
$[4]$ V. V. Borzov and E. V. Damaskinsky, Realization of the annihilation
operator for an oscillator-like system by a differential operator and
Hermite-Chihara polynomials, \emph{Integral Transforms Spec. Funct.} \textbf{13}
(2002)\newline
$[5]$ A. Odzijewicz, M. Horowski and A.
Tereszkiewicz, Integrable multi-boson systems and orthogonal polynomials,%
\emph{\ J. Phys. A: Math. Gen.} \textbf{34} (2001)\newline
$[6]$ S. T. Ali, J. P. Antoine and J. P. Gazeau, Coherent States, Wavelets,
and their Generalizations, Springer Science + Busness Media New york 1999,
2014\newline
$[7]$ W. Vogel, D. G. Welsh, Quantum optic, WILEY-VCH Verlag  \& C$_0$. KGaA,  Wheihein 2006\\
$[8]$ G. Iwata, Non-Hermitian operators and eigenfunction expansions \emph{%
Prog. Theor. Phys.} \textbf{6} (1951) \newline
$[9]$ G. N. Watson, Sc. D., F. R. S, A treatise on the theory of Bessel
Functions, Cambridge 1944\newline
$[10]$ A. O. Barut and L. Girardello, New coherent states associated with
Non-compact groups, commun. mat. phys. \textbf{21} (1971)\newline
$[11]$ W. N. Bailey, Generalized Hypergeometric Series, Cambridge Tracts in Mathematics and Mathematical Physics, Stechert-Hafner Service Agency, 1964 \newline
$[12]$ A. Voudras, K. A Penson, G. H. E. Duchamp and A. I. Solomon,
Generalized Bargmann functions, their growht an von Neumann lattice, \emph{%
J. Phys. A: Math. Theor.} \textbf{45} 244031 (2012)\newline
$[13]$ M. N. Nieto, Exact wave-function normalization constants for the $%
B\tanh z-U_0\cosh^{-2}$ and P\"{o}schl-Teller potentials, \emph{phys. Rev A,}
\textbf{17} (1978)\newline
$[14]$ P. Flagolet, M. E. H. Ismail and E. Lutwak, Classical and quantum
orthogonal polynomials in one variable, Cambridge University press 2005%
\newline
$[15]$ T. Sheecharan, P. K. Panigrahi and J. Banerji, coherent states
for exactly solvable potentials, Phys.\emph{ Rev. A.}  \textbf{69} (2004)\newline
$[16]$ H. B. Zhang, G. Y. Jiang, and S. X. Guo, Construction of
the Barut-Girardello type of coherent states for P\"{o}schl-Teller
potential, \emph{J. Math. Phys}, \textbf{55}, 122103 (2014)
\newline
$[17]$ A. M. Mathai, R. K. Saxena, Genaralized Hypergeometric function
with application in statistics and physical sciences, Springer-Verlag
Berlin. Heidelberg (1973)\newline
$[18]$ W. V. Assche, Open Problems, \emph{J. Comput. Appl. Math.}, \textbf{48} (1993)\newline
$[19]$ A. Odzijewicz, M. Horowski and A. Tereszkiewicz, Integrable
multi-boson systems and orthogonal polynomials, \emph{J. Phys. A: Math. Gen}%
. \textbf{34 } (2001)\newline
$[20]$ V. V. Borzov, Orthogonal polynomials and generalized oscillator
algebras, \emph{Integral Transforms Spec. Funct.} \textbf{12} (2001)\newline
$[21]$ R. Koekoek R. F. Swarttouw, The Askey-scheme of hypergeometric orthogonal polynomials and its q-analogue.\newline
$[22]$ N. M. Atakishiev, Quasipotential wave function of relativistic
harmonic oscillator and Pollaczeck polynomials, \emph{Teoreticheskaya
Mathematicheskaya Fizika}, \textbf{58} (1984)\newline
$[23]$ F. Pollaczek, Sur une famille de polyn\^omes orthogonaux \`{a} quatre param\`{e}tres, \emph{C. R. Acad. Sci. Paris,} \textbf{230} (1950)\\
$[24]$ V. V. Borzov and E. V. Damaskinsky, Realization of the annihilation
operator for an oscillator-like system by a differential operator and
Hermite-Chihara polynomials, \emph{Integral Transforms Spec. Funct.} \textbf{%
13} (2002) \\
$\left[25\right] $ D. Dai, W. Hu and X-S. Wang, Uniform asymptotics of
orthogonal polynomials arising from coherent states, \textit{SIGMA} \textbf{%
11} 070 (2015)\newline
$[26]$ H. Srivastava and L. Manocha, A Treatise on Generating Functions,
Ellis Horwood Ltd, London 1984\newline
$[27]$ A. P. Prudnikov, Yu. A. Brychkov, Integrals and Series volume 3
More Special Functions, Gordon and Breach Science Publishers 1990\newline
$[28]$ A. P. Prudnikov, Yu. A. Brychkov, O. I. Marichev, Integrals and Series: More special Function, Gordon and Breach Science Publishers, 1990\\
$[29]$ L. C. Andrews, Special function for Engineers and Applied
Mathematicians, Macmillan Publishing compagny, London 1985

}{
\tiny {\ \\ ${}^{\ast,\natural}$ Department of Mathematics, Faculty of
Sciences and Technics (M'Ghila), P.O.Box. 523, B\'{e}ni Mellal, Morocco. \\ \emph{ahbli.khalid@gmail.com, mouayn@gmail.com} }\\
\tiny {${}^{\flat}$ Department of Mathematics, Faculty of
Sciences, Ibn Tofail University, P.O.Box. 133, K\'{e}nitra, Morocco. \\ \emph{kayupepatrick@gmail.com} }
\end{document}